\documentclass[fleqn,useAMS,usenatbib]{mnras}
 \usepackage[T1]{fontenc}
 \usepackage{ae,aecompl}
 \usepackage{graphicx}
 
 \usepackage{multirow}
 \usepackage{color}
 \usepackage{times}
 \usepackage{amsmath}
 \usepackage{amssymb}
 
 \newcommand{\OIII}{\ion{O}{iii}}
 \newcommand{\ArIII}{\ion{Ar}{iii}}
 \newcommand{\OII}{\ion{O}{ii}}
 \newcommand{\HeII}{\ion{He}{ii}}
 \newcommand{\HeI}{\ion{He}{i}}
 \newcommand{\OI}{\ion{O}{i}}
 \newcommand{\NII}{\ion{N}{ii}}
 \newcommand{\SII}{\ion{S}{ii}}
 \newcommand{\Hb}{\mbox{H~$\beta$}}
 \newcommand{\Ha}{\mbox{H~$\alpha$}}
 \newcommand{\Hg}{\mbox{H~$\gamma$}}

\newcommand{\mbh}{$M_{\rm BH}$}

\newcommand{\s}{$\sigma$}
\newcommand{\msun}{M_{\odot}}
\newcommand{\kms}{km\,s$^{-1}$}

\begin{document}
\title[The MUSE view of QSO PG~1307+085]{The MUSE view of QSO PG~1307+085: An elliptical galaxy on the $\mathbf{M_\mathrm{BH}}$--$\mathbf{\sigma_*}$ relation interacting with its group 
environment\thanks{Based on observations made with MUSE integral field spectrograph mounted to the Yepun Very Large Telescope at ESO-Paranal Observatory under the program 60.A-9100(B)}}

\author[Husemann et al.]{B.~Husemann$^{1}$\thanks{ESO fellow, E-Mail: bhuseman@eso.org}, V.~N.~Bennert$^{2}$, J.~Scharw\"achter$^{3}$, J.-H.~Woo$^{4}$, O.~S.~Choudhury$^{5}$\\
$^{1}$European Southern Observatory (ESO), Karl-Schwarzschild-Str.2, 85748 Garching b. M\"unchen, Germany\\
$^{2}$Physics Department, California Polytechnic State University, San Luis Obispo, CA 93407, USA\\
$^{3}$LERMA, Observatoire de Paris, PSL, CNRS, Sorbonne Universit\'es, UPMC, F-75014, Paris, France\\
$^{4}$Department of Physics and Astronomy, Seoul National University, Seoul, 151-742, Korea\\
$^{5}$Leibniz-Institut f\"ur Astrophysik Potsdam, An der Sternwarte 16, 14482 Potsdam, Germany
}
\pubyear{2015}
\maketitle
\begin{abstract}
We report deep optical integral-field spectroscopy with the Multi-Unit Spectroscopic Explorer (MUSE) at the Very Large Telescope of the luminous radio-quiet quasi-stellar object (QSO) PG~1307+085 
($z=0.154$) obtained during the commissioning of the instrument. Given the high sensitivity and spatial resolution delivered by MUSE, we are able to resolve the compact ($r_\mathrm{e}\sim1.3$~arcsec) 
elliptical host galaxy. After careful spectroscopic deblending of the QSO and host galaxy emission, we infer a stellar velocity dispersion of $\sigma_*=155\pm19$~\kms. This places PG~1307+085 local 
$M_\mathrm{BH}-\sigma_*$ relation within the intrinsic scatter but offset towards a higher black hole mass with respect to the mean relation. The observations with MUSE also reveal a large 
extended narrow-line region (ENLR) around PG~1307+085 reaching out to $\sim$30~kpc. In addition, we detect a faint bridge of ionized gas towards the most massive galaxy of the galaxy group being just 
$\sim$20~arcsec (50~kpc) away. Previous long-slit spectroscopic observations missed most of these extended features due to a miss-aligned slit. The ionized gas kinematics does not show any evidence 
for gas outflows on kpc scales despite the high QSO luminosity of $L_\mathrm{bol}>10^{46}~\mathrm{erg\,s}^{-1}$. Based on the ionized  gas distribution, kinematics and metallicity we discuss the 
origin of the ENLR with respect to its group environments including minor mergers, ram-pressure stripping or filamentary gas accretion as the most likely scenarios. We conclude that PG~1307+085 is a 
normal elliptical host in terms of the scaling relations, but that the gas is most likely affected by the environment through gravity or ambient pressure. It is possible that the ongoing interaction 
with the environment, mainly seen in the ionized gas, is also be responsible for driving sufficient gas to feed the black hole at the centre of the galaxy.
\end{abstract}

\section{Introduction}
\subsection{Black hole -- spheroid relations}
An important question in the study of galaxy evolution is the origin of the relations between the mass of the central super-massive black hole (BH) and the properties of the host galaxy spheroid, 
discovered nearly two decades ago \citep{Kormendy:1995,Magorrian:1998,Gebhardt:2000,Ferrarese:2000}. Mergers or feedback from BHs growing through accretion in active galaxies have been suggested as 
potential driving forces of the implied BH-galaxy co-evolution (for a recent review see \citet{Kormendy:2013}).

Galaxies hosting active galactic nuclei (AGN) are not only promising candidates for providing direct evidence for such a co-evolution, they can also be used to study the scaling relations over cosmic 
times. Many studies suggest that BH growth precedes spheroid assembly 
\citep{Treu:2004,Woo:2006,Shields:2006,McLure:2006,Peng:2006b,Peng:2006,Treu:2007,Salviander:2007,Woo:2008,Gu:2009,Jahnke:2009,Decarli:2010,Merloni:2010,Bennert:2010b,Wang:2010,Bennert:2011}; 
however, other studies find no significant evolution with redshift \citep[e.g.,][]{Shields:2003,Shen:2008,Salviander:2013,Schramm:2013,Salviander:2015,Shen:2015,SunM:2015}.  The 
likely reason for the disagreement is that the studies are differently affected by intrinsic scatter in the relation, selection effects, and observational biases \citep[see 
e.g.,][]{Lauer:2007b,Volonteri:2011,Schulze:2014}. Another issue is that the $M_\mathrm{BH}$--$\sigma_*$ relation is not well constrained even for local AGN host galaxies at the high mass end. 
Especially for the AGN appearing as Quasi-Stellar Objects (QSOs), the bright nuclear point source often outshines its host galaxy. Measurements of the spheroid properties are therefore difficult, in 
particular the stellar velocity dispersion for which high signal-to-noise (S/N) spectra are needed. Thus, many studies focus on Seyfert galaxies often using aperture spectra which integrates over the 
central few kpc of the host galaxy \citep[e.g.,][]{Greene:2006,Woo:2006,Treu:2007,Shen:2008,Woo:2008,Matsuoka:2015}. The side effect is that, if present, a significant disc contribution may be 
included, questioning the definition of the spheroid stellar velocity dispersion in these cases. \citet{Bennert:2015} study the effect of different definitions of $\sigma_*$ in the literature for a 
sample of 66 local Seyfert-1 galaxies and find that it can vary by up to 40 per cent.

Active early-type or elliptical galaxies are therefore ideal targets to constrain the $M_\mathrm{BH}$--$\sigma_*$ relation as they lack a disturbing disc component. The colours 
of early-type QSOs, however, are still bluer than their quiescent counterparts \citep[e.g.,][]{Jahnke:2004b,Sanchez:2004b,Zakamska:2006,Trump:2013,Matsuoka:2014}. This clearly suggests that the phase 
of BH growth is also accompanied with a significant growth of the spheroid through star formation. Whether AGN feedback is responsible to quench star formation in these systems is still debated. 
Direct evidence for AGN 
feedback is provided by the kinematics of the highly-ionized gas on kpc scales, the so-called extended narrow-line region (ENLR).  In some cases, broad emission lines components with a line width of 
$>$500~\kms have been found around radio-quiet \citep[e.g.,][]{Davis:2012,Liu:2013b,Harrison:2014} and radio-loud AGN \citep[e.g.,][]{Fu:2006,Guillard:2012,Tadhunter:2014}, but not all luminous AGN 
show these prominent kinematic disturbances on kpc scales \citep[e.g.,][]{Husemann:2010,Husemann:2013a}. In some cases, the morphology and size of the ENLR rather suggests interactions with 
neighbouring galaxies \citep[e.g.,][]{Villar-Martin:2010,daSilva:2011} or the environmental gas \citep{Husemann:2011} as the origin of the gas reservoir.  

In this paper we present, for the first time, deep wide-field optical integral-field spectroscopy for the luminous radio-quiet PG~1307$+$085 at $z\sim0.15$ hosted by an undisturbed elliptical galaxy. 
The main aim of this paper is (i) to recover the optical stellar continuum and measure $\sigma_*$ as well as the systemic redshift of the host galaxy, and (ii) to characterize the physical conditions 
and kinematics of the full 2D distribution of ionized gas around PG~1307$+$085 to understand the origin of the extended gas. 

Throughout the paper we adopt canonical cosmological parameters $H_0=70~\mathrm{km}\,\mathrm{s}^{-1}\,\mathrm{Mpc}^{-1}$, $\Omega_{\mathrm{m}}=0.3$, and $\Omega_\Lambda=0.7$. 

\subsection{Previous work on PG~1307$+$085}
\begin{figure*}
 \includegraphics[width=\textwidth]{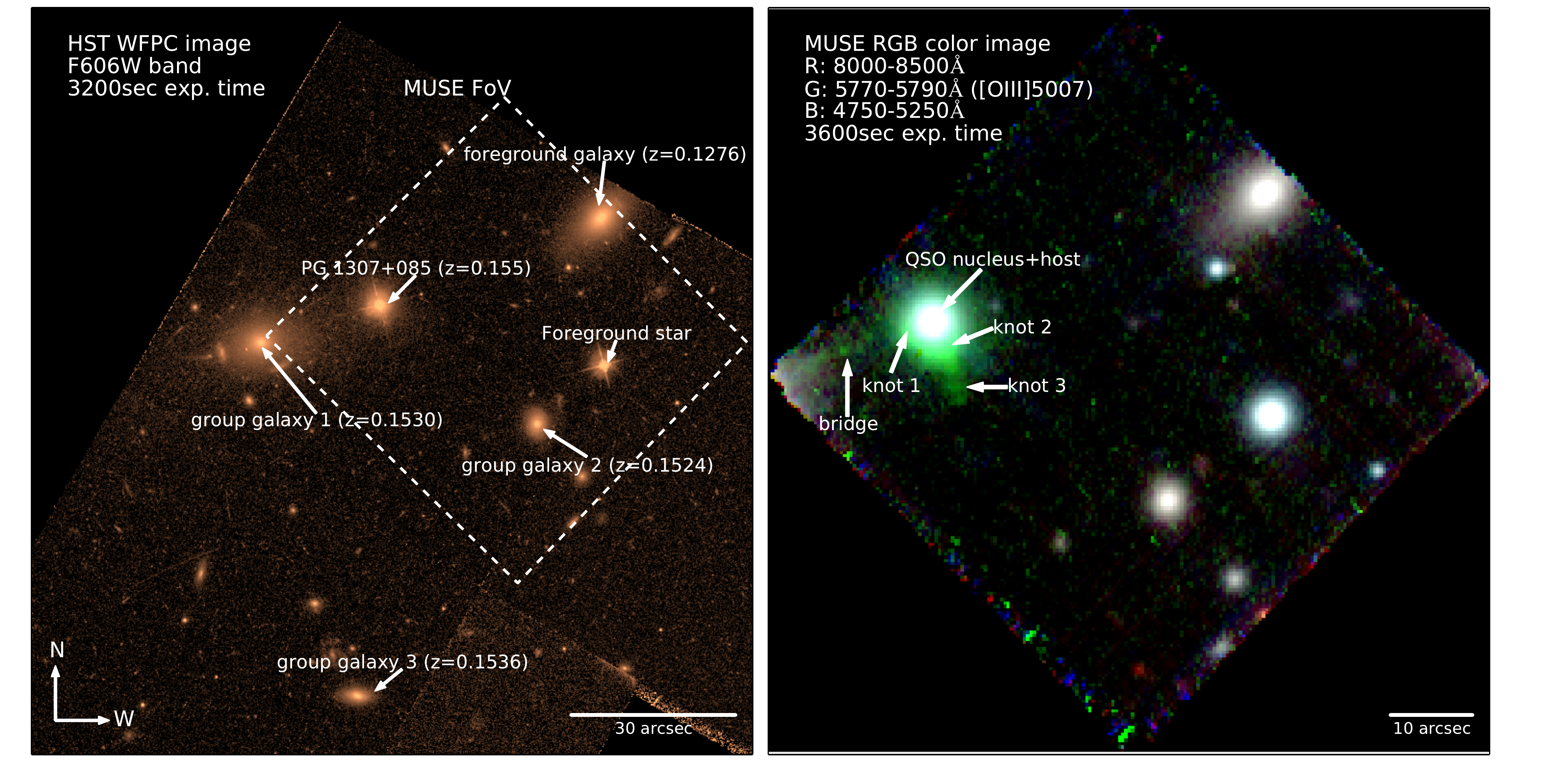}
 \caption{Single-band F606W image taken with WFPC2 aboard the \textit{HST} (left panel) and reconstructed false-colour image of the MUSE field (right panel) targeting the QSO PG~1307$+$085. Several 
galaxies are part of rich galaxy group extending beyond the \textit{HST} image that were spectroscopically identified by \citet{Prochaska:2011} are labelled on the \textit{HST} image. The centre 
of the 
group is most likely the luminous ($\sim 4L_*$) elliptical galaxy to the south-east of PG~1307$+$085. For the MUSE false-colour image we choose two broad continuum bands and a narrow band centred on 
the redshifted [\OIII]$\,\lambda5007$ emission line to highlight the ionized gas distribution around the QSO. Besides the bright [\OIII] emission of the nucleus we discovered an extended 
emission-line region with a size up to $\sim$10~arcsec from the nucleus. The three main structures of the extended emission-line regions are two bright knots to the east and south, respectively, and 
a low surface brightness tail extending further south of knot 2, which are labelled in the figure accordingly.}
 \label{fig:overview}
\end{figure*}
The QSO PG~1307$+$085 was discovered by the Palomar-Green (PG) survey as a UV-excess source at the coordinates $13^\mathrm{h}09^\mathrm{m}47^\mathrm{s}~+08^\circ19'48''$ (J2000). A magnitude of 
$m_B=15.28$ and a redshift of $z=0.155$  was 
reported by the Bright Quasar Survey \citep[(BQS)][]{Schmidt:1983}. Very Large Array (VLA) observations of the BQS sample by \citet{Kellermann:1989} detected only weak radio emission 
($S_{1.4\mathrm{GHz}}=3.5~\mathrm{mJy}$), consequently classifying PG 1307+085 as a radio-quiet QSO with an R parameter of 0.1  following the definition and classification of 
\citet{Kellermann:1989}. As a bright low-redshift QSO, PG~1307$+$085 has been studied in detail with follow-up observations from X-ray to radio wavelength.

The underlying host galaxy was first detected by \citet{McLeod:1994} in the $H$ band from the ground. They reported a host magnitude of $m_H=15.25$~mag, a factor of four fainter than the QSO. 
Observations with the \textit{Hubble Space Telescope (HST)}  \citep{Bahcall:1994,Veilleux:2009,Bentz:2009a}  and deep adaptive-optics ground-based $H$ band imaging \citep{Guyon:2006} spatially 
resolved 
the underlying host galaxy and consistently classified it as an undisturbed elliptical galaxy.  From 2D image modelling, \citet{Bahcall:1997} and \citet{Veilleux:2009} reported  an 
effective radius of $r_\mathrm{e}\sim1.3$~arcsec (3.5~kpc) and a host galaxy magnitude of $m_{\mathrm{F606W}}=17.8$~mag and $m_{H}=15.21$~mag, respectively, whereas \citet{Guyon:2006} inferred a 
 slightly larger ($r_\mathrm{e}=4.7$~arcsec) and brighter ($m_{H}=14.71$~mag) host galaxy. \textit{ Spitzer} and \textit{IRAS} satellite observations reveal that PG~1307$+$085 is far-infrared faint 
with  
$L(60~\umu\mathrm{m})/L(24~\umu\mathrm{m})<1$  and weak emission from polycyclic aromatic hydrocarbons \citep{Schweitzer:2006,Netzer:2007}. This implies a significantly lower star formation rate 
compared to other QSOs of similar luminosity. 

PG~1307$+$085 is found to reside in a group environment, based on a spectroscopic survey of galaxies around nearby QSOs \citep{Prochaska:2011}. The survey found ten galaxies brighter than 
$m_R<19.5$~mag with a redshift difference of $\Delta z<0.005$ closer than $<$5~arcmin ($<$780~kpc) from the QSO. The brightest galaxy of that group with $m_R=16.77$~mag is just $\sim$21~arcsec 
($\sim$60~kpc) away from the QSO. With an absolute magnitude of $M_R=-22.56$~mag, the galaxy is a massive $4L_*$ galaxy. 

The ionized gas around a large sample of QSOs including PG~1307$+$085 has initially been studied by \citet{Stockton:1987}. They detected no ENLR around this QSOs in  
ground-based narrow-band images centred on the [\OIII]$\lambda5007$ line. Narrow-band images with the \textit{HST} revealed the first hints of an ENLR on scales of 
1.4~arcsec \citep{Bennert:2002}. Deeper observations with \textit{HST} centred on the [\OII], \Hb, [\OIII] and Pa~$\alpha$ line basically agree in the apparent extension of ionized gas 
\citep{Young:2014}. Based on the line ratio diagnostics the authors argue that the gas is mainly ionized by ongoing star formation with a SFR of about $\sim20M_\odot\,\mathrm{yr}^{-1}$. However, 
\textit{HST} narrow-band images are often too shallow to capture the low surface 
brightness part of the large-scale ENLRs \citep[e.g.][]{Husemann:2013a}. Optical long-slit spectroscopy with EMMI at the New Technology Telescope (NTT) \citep{Leipski:2006a} and FORS1 at the Very 
Large Telescope (VLT) \citep{Oh:2013} reveal indeed a spatially extended [\OIII]$\lambda5007$ line out to 4~arcsec from the nucleus. The slits for both observation were taken at the same position 
angle 
and consistently show a spatially asymmetric distribution of a narrow line component with a dispersion of $\sim70$~\kms. A distinct broader emission-line component has also been 
reported by both studies, but it remained unclear whether this originates from the QSO due to the wings of the point-spread function (PSF) or true jet-cloud interaction with a candidate 
low-luminosity 
radio jet \citep{Leipski:2006b}.

\section{Observations and data reduction}
The QSO PG~1307+085 was observed with the Multi-Unit Spectroscopic Explorer \citep[MUSE,][]{Bacon:2010,Bacon:2014a} at the VLT on 2014 May 6 as part of the instrument commissioning. 
MUSE is a novel optical integral-field spectrograph mounted to the Nasmyth focus of the UT4 telescope Yepun. In the wide-field mode, MUSE captures a contiguous $1\arcmin\times1\arcmin$ 
field-of-view (FoV) with a sampling of 0.2~arcsec. The covered nominal wavelength range is fixed from 4750\AA\ to 9350\AA\ at a spectral resolution of about $\sim$2.5\AA.
The corresponding data set became public in August 2014. The total integration time of 3600~s on source was split into six 600~s long exposure. A small dither and a 
90$^{\circ}$ field rotation was applied between individual exposures to aid removal of cosmics and artefacts as well as to suppress flat-fielding residuals during the final combination of the 
frames. Observations were performed under good ambient conditions with $<30$ per cent of moon illumination, clear sky and a seeing ranging between 0.8--1.1~arcsec. 

Data reduction is carried out using the MUSE pipeline (version 0.18.2) which performs basic reduction steps including bias subtraction, flat fielding, wavelength calibration and geometric 
calibration of the MUSE field \citep[][Weilbacher et al. in prep.]{Weilbacher:2012}. A master calibration generated during the commissioning run is applied for the geometric solution of the 
instrument field. All six exposure are combined to create the final datacube with the pipeline using the drizzle resampling method. During this step, the data is also flux calibrated and the sky 
is subtracted based on the empty regions within the large 1~arcmin FoV of MUSE. 

We notice that the sky subtraction of the given pipeline version is not optimal and leaves strong sky line residuals in the data. The residuals appear of systematic nature so that we apply an 
additional correction. We create an average residual sky spectrum from a $5\times5$~arcsec${}^2$ region close to the QSO, but without any signal from the QSO or ENLR itself. Then 
we subtract the residual sky spectrum from the data which significantly improves the sky subtraction over large parts of the MUSE FoV. In addition, we estimate and correct for telluric absorption 
bands based on the spectrum of relatively bright star in the field. The telluric absorption correction is particular important for PG~1307+085 because the broad \Ha\ line and  the important 
[\NII]$\lambda6583$ line are redshifted into one of the strongest absorption bands. Finally, we correct the spectra for Galactic dust attenuation adopting the Milky-way dust attenuation law by 
\citet{ODonnell:1994} adopting an extinction of $A_V=0.095$~mag along the line-of-sight \citep{Schlafly:2011} and assuming $R_V=3.1$.

Given that the seeing during the observation was only $>$0.8~arcsec, we binned the final reduced cube by $2\times2$ pixels leading to a sampling of 0.4~arcsec. The binning increases the S/N 
per spatial pixel while the data remain Nyquist sampled with respect to the intrinsic spatial resolution. Our binned MUSE data cube is publicly 
available\footnote{http://www.bhusemann-astro.org/datasets/}.

\section{Analysis}
\subsection{Reconstructed narrow-band and broad-band images}
We extract a $20$\AA\ wide narrow-band image centred on the redshifted [\OIII]$\lambda5007$ line and two broad band images covering the 4750--5250\AA\ and 8000-8500\AA\ pass 
bands, respectively. The corresponding pseudo RGB image of the field is shown in Fig.~\ref{fig:overview} (right panel) compared to an archival \textit{HST} image taken with WFPC2 camera in the 
F606W 
filter (left panel). While the QSO host galaxy has been classified as a round elliptical without any signs of distortion, the MUSE data reveal an extended and complex distribution of ionized gas 
around the QSO. We identify three prominent structures across the ENLR: (1) a bright knot south-east of the nucleus, (2) a bright knot plus shell-like structure south of the nucleus, and (3) a faint 
knot about 10~arcsec (26~kpc) south from the nucleus. In addition to PG~1307$+$085, a few galaxies of the group are also covered by the MUSE FoV. The bright elliptical galaxy is only partially 
covered at the east edge of the FoV, but we are able to detect a very low surface brightness bridge of ionized gas between this galaxy and the QSO host galaxy.

\begin{figure}
\centering
 \includegraphics[width=0.35\textwidth]{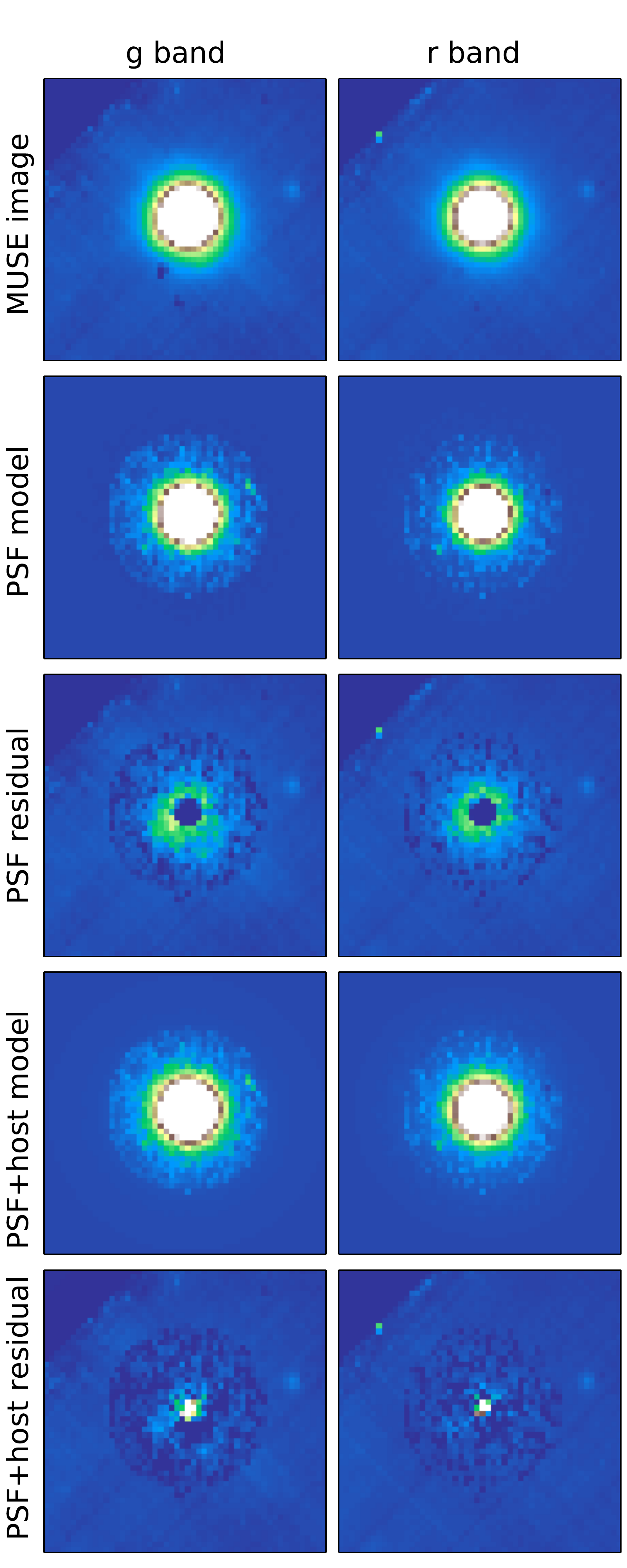}
 \caption{Rest-frame $g$ and $r$ broad-band images and corresponding PSF and PSF+host galaxy models with residuals. All images are displayed with a common linear scaling. Despite the 
flat-fielding residuals, the light is significantly more extended compared to a pure point source. Parameters from the best-fitting PSF+host model are given in Table~\ref{tbl:host_decomp}.}
 \label{fig:host_decomp}
\end{figure}

\subsection{QSO-host deblending}
Even at good seeing conditions, the bright QSO significantly contaminates a large area due to the wings of the PSF and the high contrast ratio between unresolved and extended 
emission. Here we take advantage of the broad Balmer lines of the QSO to measure the PSF \citep[e.g.][]{Jahnke:2004}. Our iterative algorithm to deblend the point-like 
emission of the QSO and the extended host galaxy emission is described in detail in \citet{Husemann:2013a,Husemann:2014}, we here just briefly outline the process. 

\begin{figure*}
 \includegraphics[width=\textwidth]{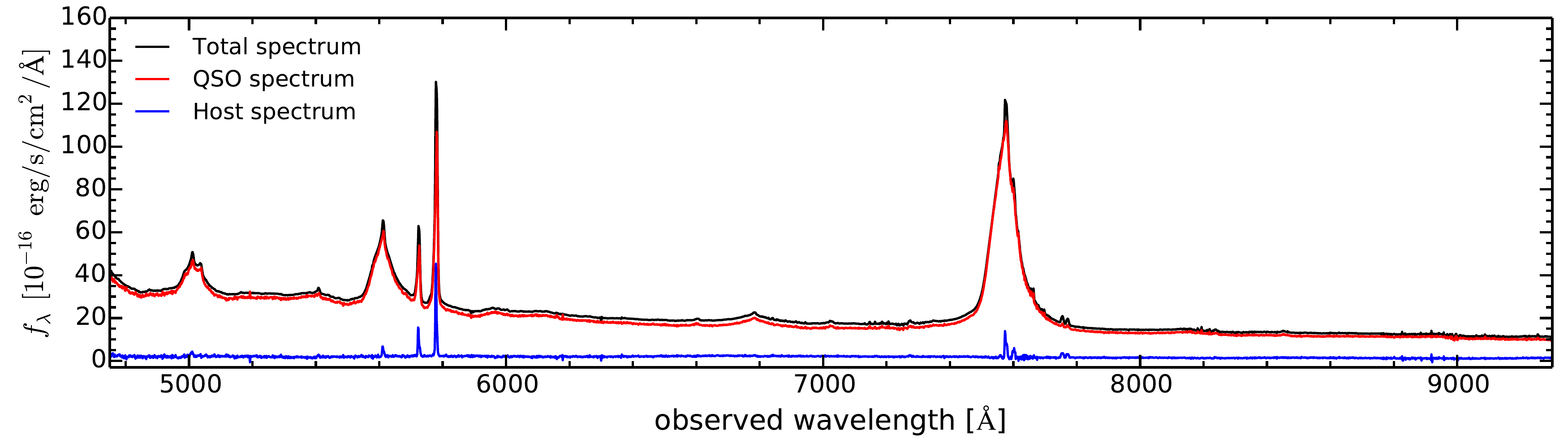}
 \caption{The total, pure QSO and the host galaxy spectrum containing spatially resolved emission within an aperture of 3.2~arcsec radius. The QSO exceeds the host galaxy light by a factor $\sim$10 
at 
5250\AA\ (rest frame). No significant broad emission-line residuals appear in the host galaxy spectra, but strong narrow-emission lines are clearly present from the surrounding kpc-scale ENLR as seen 
in the narrow-band image shown in Fig.~\ref{fig:overview}.}
 \label{fig:decomp_spectra}
\end{figure*}

First, we measure a PSF from broad emission lines by mapping the intensity in the line wings above the adjacent continuum flux. Since the PSF is smoothly changing as a function of wavelength, we 
determine the PSF from the broad \Hg, \Hb, \HeI, and \Ha\ lines. Then we reconstruct a PSF cube by normalizing all PSFs to one at the brightest pixel and interpolate the intensity 
maps with a two-order polynomial function along wavelength. An almost pure QSO cube is produced by multiplying the normalized PSF cube with QSO spectrum from the brightest spaxel\footnote{denotes a 
single spatial pixel in the datacube and consists of an entire spectrum}. However, there is some host galaxy contamination even in the QSO spectrum that we need to subtract to avoid significant 
over-subtraction. Since the radial surface brightness distribution of the host galaxy is necessarily more extended than the PSF, we co-add the residual spectra around the brightest spaxel after the 
QSO subtraction to obtain a best guess on the host galaxy spectrum. We re-scale this host galaxy spectrum in total flux so that it matches the \emph{expected} host galaxy surface brightness at that 
brightest spaxel. In an iterative scheme we subsequently subtract the re-scaled host spectrum from the original QSO spectrum in each iteration to clean it from the host contamination. The process 
converges after a few iterations and we stop after ten in this particular case.
\begin{table}
\centering
\caption{Best-fitting 2D image decomposition parameters}
\label{tbl:host_decomp}
 \begin{tabular}{lll}\hline
\textbf{Parameter} & \textbf{$g$ band} & \textbf{$r$ band} \\\hline
 QSO 	   & 15.5~mag    & 15.2~mag   \\
 host      & 17.9~mag    &  17.4~mag  \\
 $n$       & 4 (fixed)    & 4 (fixed) \\
 $r_\mathrm{e}$     &  1.3~arcsec (3.5~kpc)  & 1.2~arcsec (3.2~kpc) \\ 
 $b/a$     &   0.98       &  0.97 \\
 PA        &   $67^{\circ}$ &  $64^{\circ}$\\\hline
 \end{tabular}
\end{table}
To estimate the radial surface brightness profile of the host galaxy, we extracted rest-frame $g$ and $r$-band images from the initial data as well as the corresponding PSFs from the 
PSF cube.  Then we used \textsc{galfit} \citep{Peng:2002,Peng:2010} to model the 2D surface brightness as a superposition of a point source and a Sersi\'c profile. We fixed the Sersi\'c 
index to $n=4$ because the host galaxy is known to be an elliptical galaxy from higher resolution \textit{HST} observations \citep{Bahcall:1994,Veilleux:2009}. In Fig.~\ref{fig:host_decomp} we show 
the 
model and residual of a PSF model only  and a PSF+Sersi\'c model for the $g$ band (left panel) and  $r$ band (right panel) images, respectively. Our best-fitting parameters 
for both bands are given in Table~\ref{tbl:host_decomp}. 

MUSE is not designed to be a broad-band imager and the systematic uncertainty of the flat-fielding across the field is strongly limited by the preliminary calibration plan during the commissioning 
run. Thus, it is not possible to obtain very precise measurements on the host galaxy morphological parameter. However, the residuals of the PSF-only model are significant and highlight that the 
underlying host galaxy of PG~1307+085 is well resolved, even in the seeing-limited MUSE data. The model for the $g$ and $r$ bands leads to an effective radius of $\sim$1.3~arcsec which is 
in perfect agreement with the measurements with \textit{HST} reported by \citet{Bahcall:1997} and \citet{Veilleux:2009}. 

After combining the information on the host galaxy 2D profile with the wavelength-dependent PSF measurements, we successfully decompose the spatially unresolved QSO emission from 
the spatially resolved emission. We show the original and decomposed spectra within a radius of 16~pixels (3.2~arcsec) around the QSO in Fig.~\ref{fig:decomp_spectra}. The QSO clearly dominates 
the continuum emission by a factor of $\sim$10 at 5250\AA\ (rest frame) close to the wavelength of the \ion{Mg}{i}b\,$\lambda$5175\AA\ stellar absorption line. We recover prominent narrow emission 
lines 
even in this circumnuclear region which we already know extends much further as seen in the narrow-band image (Fig.~\ref{fig:overview}) even before applying the QSO-host debelending. 

\subsection{Stellar continuum and velocity dispersion}
\begin{figure}
\centering
 \resizebox{\hsize}{!}{\includegraphics{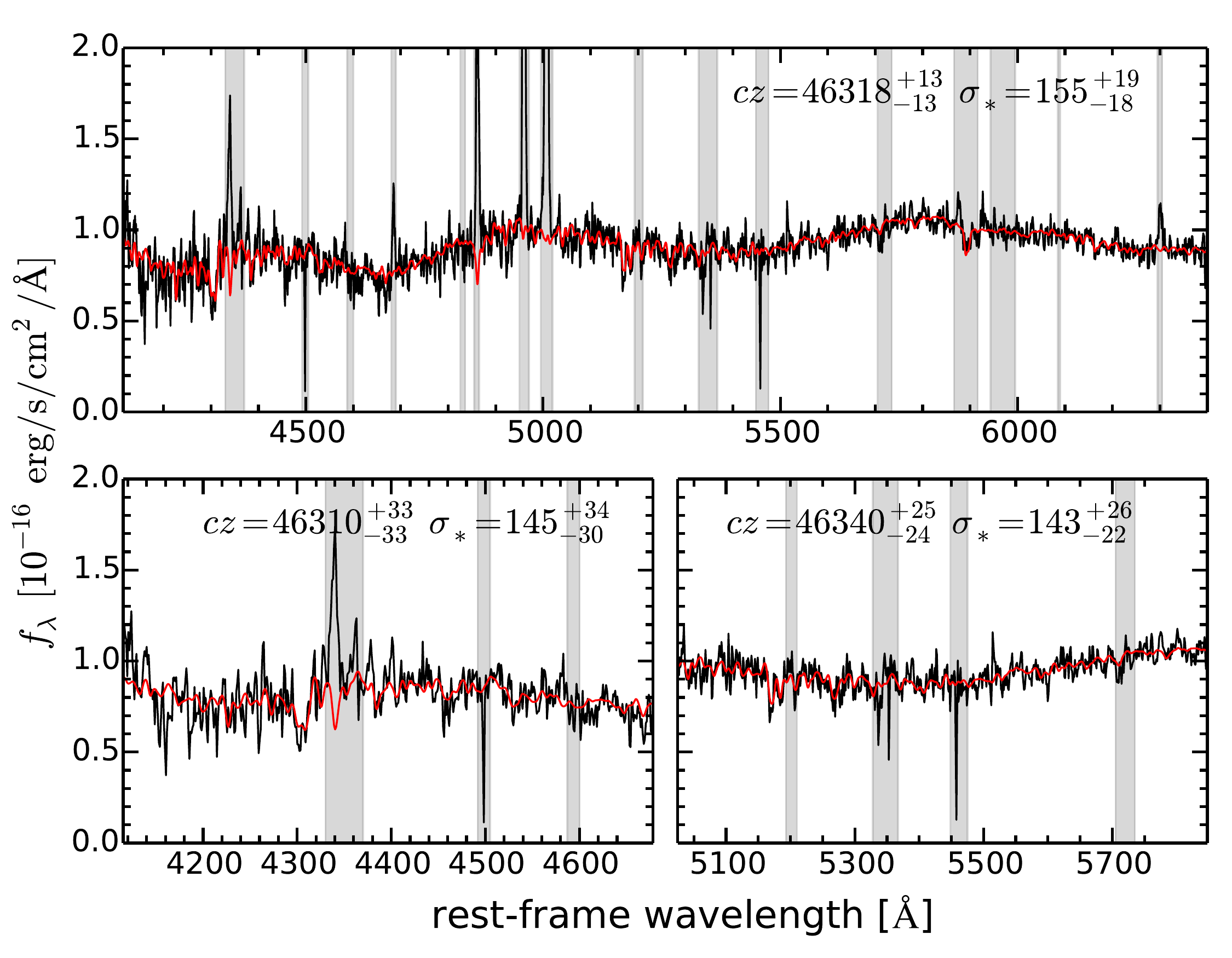}}
 \caption{Host galaxy continuum spectrum within an aperture of 2~arcsec in radius. The spectrum is slightly smoothed from a spectral resolution of 2.5 to 3.0~\AA\ with a Gaussian kernel. We 
model the spectrum as a super-position of single-stellar population spectra \citep{Vazdekis:2010} generated from the MILES stellar library \citep{Sanchez-Blazquez:2006}, adopting a Gaussian kernel 
for 
the LOSVD which is shown as the red line. The modelling was performed independently on the entire spectrum up to the wavelength of \Ha\ (upper panel),  covering a wavelength region centred around the 
$G$ band stellar absorption feature (lower left panel) and the \ion{Mg}{i}b+Fe stellar absorption features (lower right panel). All measurements agree within the error bars, but the full spectral 
fitting is considered more robust as template mismatches are minimized. }
 \label{fig:host_spec_fit}
\end{figure}

\begin{figure*}
 \includegraphics[width=\textwidth]{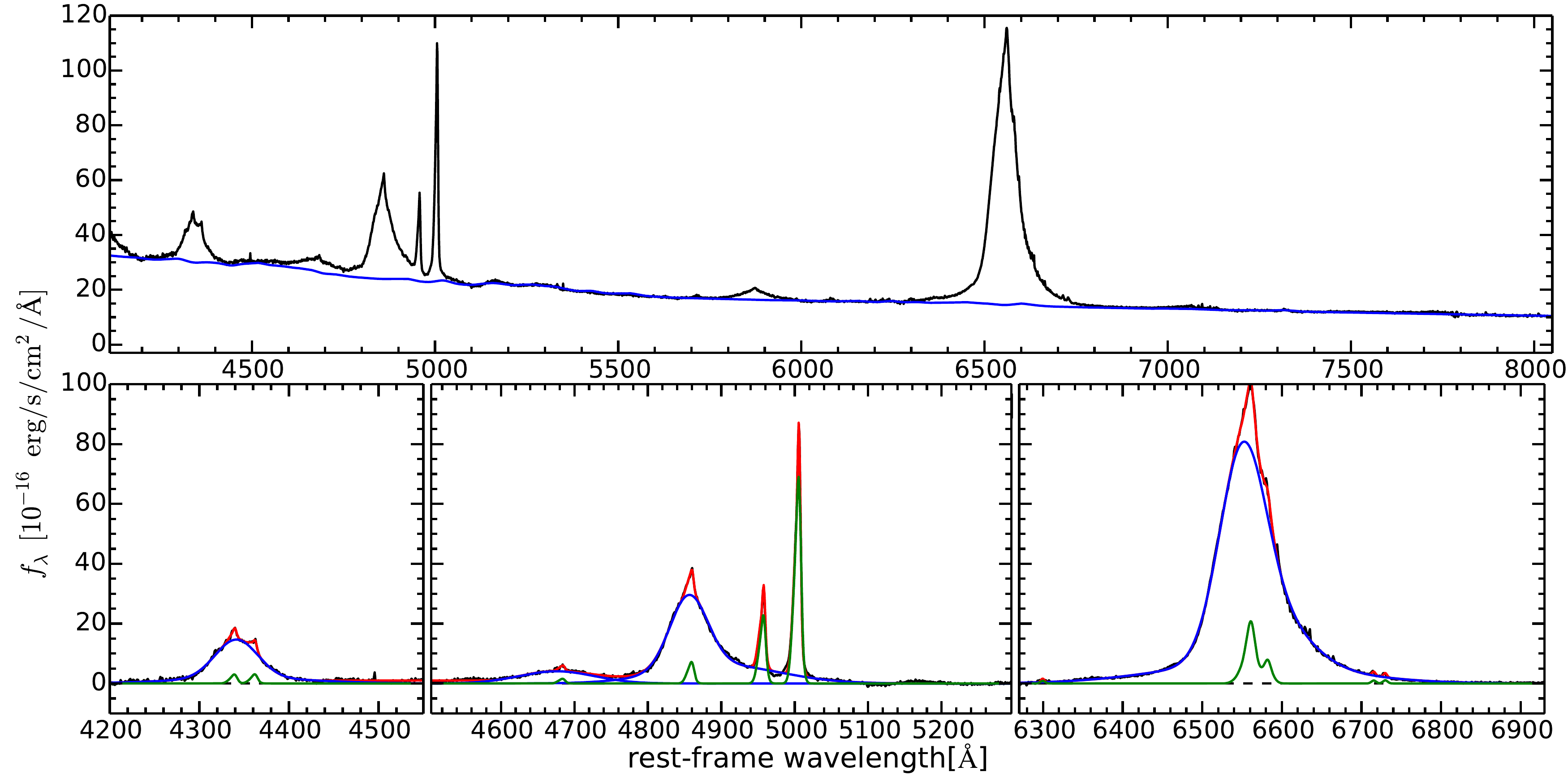}
 \caption{\textit{Upper panel:} Integrated QSO spectrum for PG1307$+$085 after the debelending process. The reconstructed pseudo-continuum is shown as blue line which is a superposition of a 
power-law, with two different slopes blue-ward and red-ward of 5650\AA\ (rest frame), and the smoothed Fe\,II template from \citet{Boroson:1992}. \textit{Lower panels:} Emission-line spectra after 
continuum subtraction centred on the broad \Hg\ (left panel), \Hb\ (middle panel) and the \Ha\ line (right panel). The best-fitting model is shown as red solid line which are composed 
of several Gaussians for the broad-line components (blue line) and the spectrum of all narrow lines (green line). See text for details about the 
emission-line model, parameters and constraints.}
\label{fig:QSO_fit}
\end{figure*}
In Fig.~\ref{fig:host_spec_fit} we show the host galaxy spectrum extracted from an aperture of 2~arcsec in radius to capture all the light within the effective radius given a 1~arcsec seeing. We 
slightly degrade the spectral resolution from 2.5\AA\ Full Width Half Maximum (FWHM), estimated from the width of the sky lines, to 3.0\AA\ FWHM. This allows us to model the spectrum  with stellar 
population synthesis spectra from the Medium-resolution Isaac newton telescope Library of Empirical Spectra (MILES) \citep{Sanchez-Blazquez:2006}. The intrinsic rest-frame spectral resolution of the 
spectra is 2.51\AA\ \citep{Falcon-Barroso:2011} corresponding to 2.89\AA\ at the redshift of the QSO. The slight spectral smoothing also increased the S/N of the spectrum from S/N$\sim$8 to 
S/N$\sim$10 per pixel. Strong stellar absorption features are visible in the spectrum in particular the prominent $G$ band at 4300\AA, \ion{Mg}{i}b triplet at 5175\AA, various adjacent Fe 
lines as well as the NaD line at 5892\AA. The continuum shows a small low frequency spectral shape deviation due to systematic uncertainties in the QSO host deblending. Those systematics are caused 
by 
(1) uncertainties to correct the positional shifts due to differential atmospheric refraction, and (2) the uncertainties in the broad-line PSF measurements and subsequent interpolation with 
wavelength. 

Here, we use our own software {\tt PyParadise} (Husemann, Choudhury \& Walcher in prep.) which is an extended Python version of {\tt paradise} \citep{Walcher:2015} to perform stellar population 
spectral synthesis to fit the stellar continuum. The algorithm is insensitive to the global continuum shape as it first divides the spectrum and all template spectra by a running mean after linearly 
interpolating masked wavelength regions. A detailed description of the algorithm is presented in \citet{Walcher:2015} which we summarize below. The main improvements for {\tt PyParadise} 
are that (1) a Markov-Chain-Monte-Carlo (MCMC) algorithm is used to find the parameters and associated errors for the line-of-sight velocity distribution (LOSVD), and (2) that only the high S/N 
template library spectra are resampled during the fitting while the input spectrum remains unchanged to preserve its noise properties.

In the initial iteration, a single stellar population spectrum is drawn from the library and the best-fitting LOSVD, in terms of redshift $z$ and velocity dispersion $\sigma_*$, is 
estimated by means of a MCMC algorithm \citep[PyMC,][]{Patil:2010} in which the spectrum is smoothed with corresponding Gaussian kernels. Here, we parametrize the LOSVD only by a simple 
Gaussian and avoid to model higher moments given the relatively low S/N of the stellar continuum. In order to match the spectral resolution of the data and the template spectra, we smooth the 
template spectra, after redshifting to the initial redshift guess $z_\mathrm{guess}$, to match the spectral resolution of the data. For the priors we assume a flat probability 
distribution within $\|c(z-z_{\mathrm{guess}})\|<500$~\kms\ and $50$~\kms$<\sigma_*<$400~\kms. From the posterior distribution we select the most likely LOSVD and apply it to all normalized spectra 
of 
the template library and use a non-negative linear least square algorithm to determine the best-fitting linear coefficients. In all following iterations, this two step process is repeated but the 
initial spectrum is replaced by a linear combination of the template library based on best-fitting linear coefficients of the previous iteration without applying the LOSVD. The LOSVD parameters and 
linear coefficient of the templates converge quickly and here we stop after five iterations. 

The best-fitting spectra are shown in Fig.~\ref{fig:host_spec_fit} based on the library of single stellar population models presented by \citet{Vazdekis:2010} from the MILES stellar library for a 
universal Kroupa initial mass function. We perform an independent modelling of three different wavelength regions; (i) the entire spectrum up to \Ha\ (upper panel), (ii) the wavelength region 
centred on the $G$ band (lower left panel) and (iii) the wavelength region centred on \ion{Mg}{i}b+Fe absorption lines (lower right panel). The best-fitting parameters for the $z$ and $\sigma_*$ 
agree within the $1\sigma$ error bars inferred from the almost symmetric posterior distribution as the 16\% and 83\% percentiles. We consider the model of the full wavelength range to be 
more 
robust as the simultaneous fitting of significantly more feature reduces the systematic errors caused by template mismatch. Our best-fitting parameters for the LOSVD are 
$cz=46318_{-13}^{+13}$~\kms\ ($z=0.1544\pm0.0001$) and $\sigma_*=155_{-18}^{+19}$~\kms.  Thus, we obtain a very reliable measurement of the systemic redshift which is import to interpret the ionized 
gas kinematics.

\subsection{QSO spectrum and BH mass estimate}
The QSO spectrum is shown in Fig.~\ref{fig:QSO_fit} after subtracting the host galaxy contribution. We model the spectrum as a super-position of a continuum and emission lines components following 
the 
scheme of \citet{Shen:2008b}. We assume a power-law function as the primary continuum with two different slopes in the blue and red part of the rest-frame optical spectrum 
\citep[e.g.][]{VandenBerk:2001}. Due to the various broad lines and Fe\,II bands we can only infer the continuum flux at three rest-frame wavelengths: 
4200\AA, 5650\AA, and 7400\AA. Based on the best-fitting power-law  we infer a continuum flux density at 5100$\AA$ (rest-frame) of $f_{5100}=21.5\times 
10^{-16}~\mathrm{erg}\,\mathrm{s}^{-1}\,\mathrm{cm}^{-2}\,\mathrm{\AA}^{-1}$ which corresponds to a continuum luminosity of $L_{5100}=8\times10^{44}~\mathrm{erg}\,\mathrm{s}^{-1}$. 
An additional pseudo-continuum is produced by the FeII bands that we model with the broadened FeII template spectrum of I~Zw~1 provided by \citet{Boroson:1992}. The width of the smoothing 
kernel is empirically constrained to minimize the residuals for the prominent FeII band at 5100\AA--5450\AA.  

We model all the emission lines above the continuum with  multiple Gaussian components (see Fig.~\ref{fig:QSO_fit}). Two broad Gaussian components are necessary 
to model the broad \Hg\ and \Hb\ and \Ha\ line shape, respectively, whereas one broad component is sufficient for the broad \HeII\,$\lambda$4685\AA\ line. Prominent narrow lines are 
also seen in \Hg, [\OIII]\,$\lambda$4363\AA, \HeII\,$\lambda$4685\AA, \Hb, [\OIII]\,$\lambda\lambda4960,5007$, [\OI] $\lambda6300$\AA, \Ha, [\NII]\,$\lambda\lambda6548,6583$\AA, and 
[\SII]\,$\lambda\lambda6717,6730$\AA. All of the narrow lines except [\OI] and [\SII] appear asymmetric with a blue wing, so that we model those narrow lines with two Gaussian components. Such a 
broad asymmetric wing is a common feature in the spectra of luminous QSOs \citep[e.g.][]{Mullaney:2013} and considered to be a genuine signature of AGN-driven outflows at least in 
the circumnuclear region.  The best-fitting parameters for these two narrow-line components are listed in Table~\ref{tbl:line_measurements}. In contrast to the broad lines, we force the redshift and 
intrinsic velocity dispersion of the two narrow-line kinematic components to be the same. This greatly reduces the number of free parameters and leads to a more robust fit.

The broader component of the narrow lines has a velocity dispersion of about $\sim$300~\kms\ compared to $\sim$100~\kms\ for the other components. The broader component 
is blue-shifted by about 60~\kms. The velocity dispersion of the bluer component and its blue-shift with respect to the narrower component are smaller than the average value of 
luminous AGN as inferred by \citet{Mullaney:2013}. We discuss the physical conditions of this ionized gas together with the extended emission in the next section. One difficulty 
in their interpretation is that the fluxes of the narrow \Hb\ and \Ha\ line are unreliable due to the systematic error in the model of the complex broad Balmer line shapes that are 
compensated by a skewed flux ratio of the two narrow line components.

We measure a broad \Hb\ line width of $\mathrm{FWHM}_{\mathrm{H}\beta}=4250\pm30$~\kms and $\sigma_{\mathrm{H}\beta}=2347\pm21$~\kms. Those errors correspond only to
the random uncertainties given the S/N of the spectrum, but not the systematic uncertainties due to line blending. The FWHM is in good agreement with the measurements of \citet{Kaspi:2000} and 
\citet{Shen:2011}, but 1000~\kms\ narrower than reported by \citet{Marziani:2003} and \citet{Peterson:2004}. \citet{Oh:2013} reported a line dispersion of 
$\sigma_{\mathrm{H}\beta}=1885\pm10$~\kms\ which is 460~\kms\ narrower than our measurements.
It is unclear whether this is attributed to time variability in the line shape or systematic uncertainties of the \Hb\ deblending into a narrow and broad component since \citet{Kaspi:2000} and 
\citet{Peterson:2004} obtain very different results based on the same spectra. 

A reverberation time-lag of $\tau=105_{-36}^{+46}$ light days for PG~1307+085 was initially reported by \citet{Peterson:2004} for the broad \Hb\ line, but significantly revised to 
$\tau=188_{-3.7}^{+5.3}$ light days by \citet{Zu:2011}. The corresponding virial product 
$VP=c\tau\sigma_\mathrm{line}^2/G=120_{-14}^{+21}\times10^{6}\msun$ for the broad \Hb\ line dispersion as measured by \citet{Peterson:2004} combined with a virial factor of 
$f=4.47_{-1.2}^{+1.4}$ \citep{Woo:2015} leads to a reverberation-based BH mass of $M_\mathrm{BH}=5.3_{-1.9}^{+2.0}\times10^{8}\msun$.
Since we infer a broader \Hb\ line dispersion from the analysis of the MUSE spectrum, we also compute a single-epoch virial BH mass replacing the reverberation-mapped BLR size with the 
size-luminosity relation of \citet{Bentz:2013}. This yields a BH mass of $M_\mathrm{BH}=4.9\times10^{8}\msun$ which is well within the uncertainties of the reverberation-based BH mass. We therefore 
refer to the reverberation-mapped BH throughout the paper.

Assuming a mean bolometric correction factor of ten for the continuum luminosity at 5100\AA\ following \citet{Richards:2006}, we obtain a bolometric luminosity of 
$\log(L_\mathrm{bol}/[\mathrm{erg\,s}^{-1}])=45.9$. This corresponds to an Eddington ratio of $\lambda=L_\mathrm{bol}/L_\mathrm{Edd}=0.3$ for PG~1307+085. 
\begin{figure}
 \resizebox{\hsize}{!}{\includegraphics{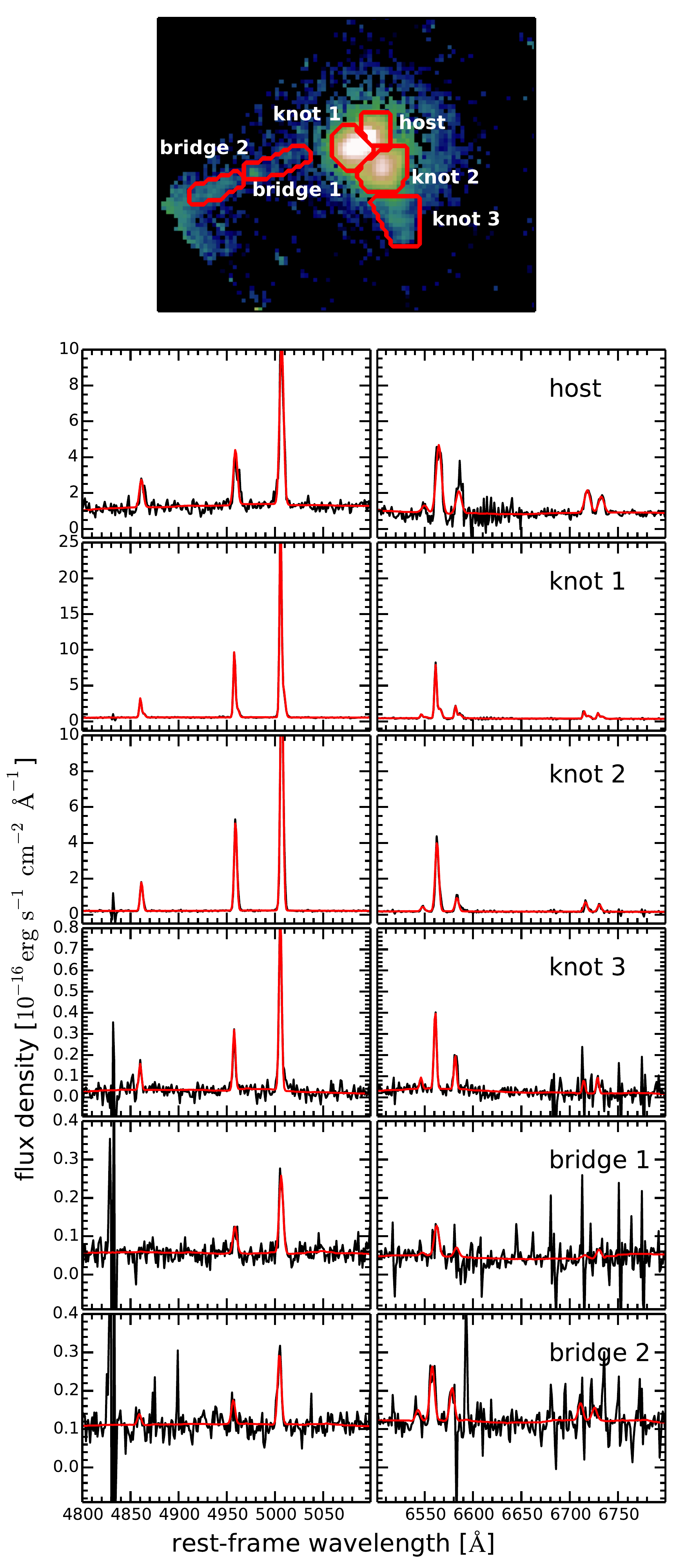}}
 \caption{Co-added spectra and best-fitting emission-line models for several prominent emission-line regions indicated in the narrow-band [\OIII] image on the top. Two spectral windows covering the 
\Hb\ and [\OIII]\,$\lambda\lambda4960,5007$ (left panels) and the \Ha, [\NII]\,$\lambda\lambda6548,6583$, and [\SII]\,$\lambda\lambda6717,6730$ (right panels) are shown. The two ''bridge'' 
spectra suffer more strongly from the sky line residuals given the low surface brightness of the emission lines.}
 \label{fig:spectra}
\end{figure}
\begin{figure*}
 \includegraphics[width=\textwidth]{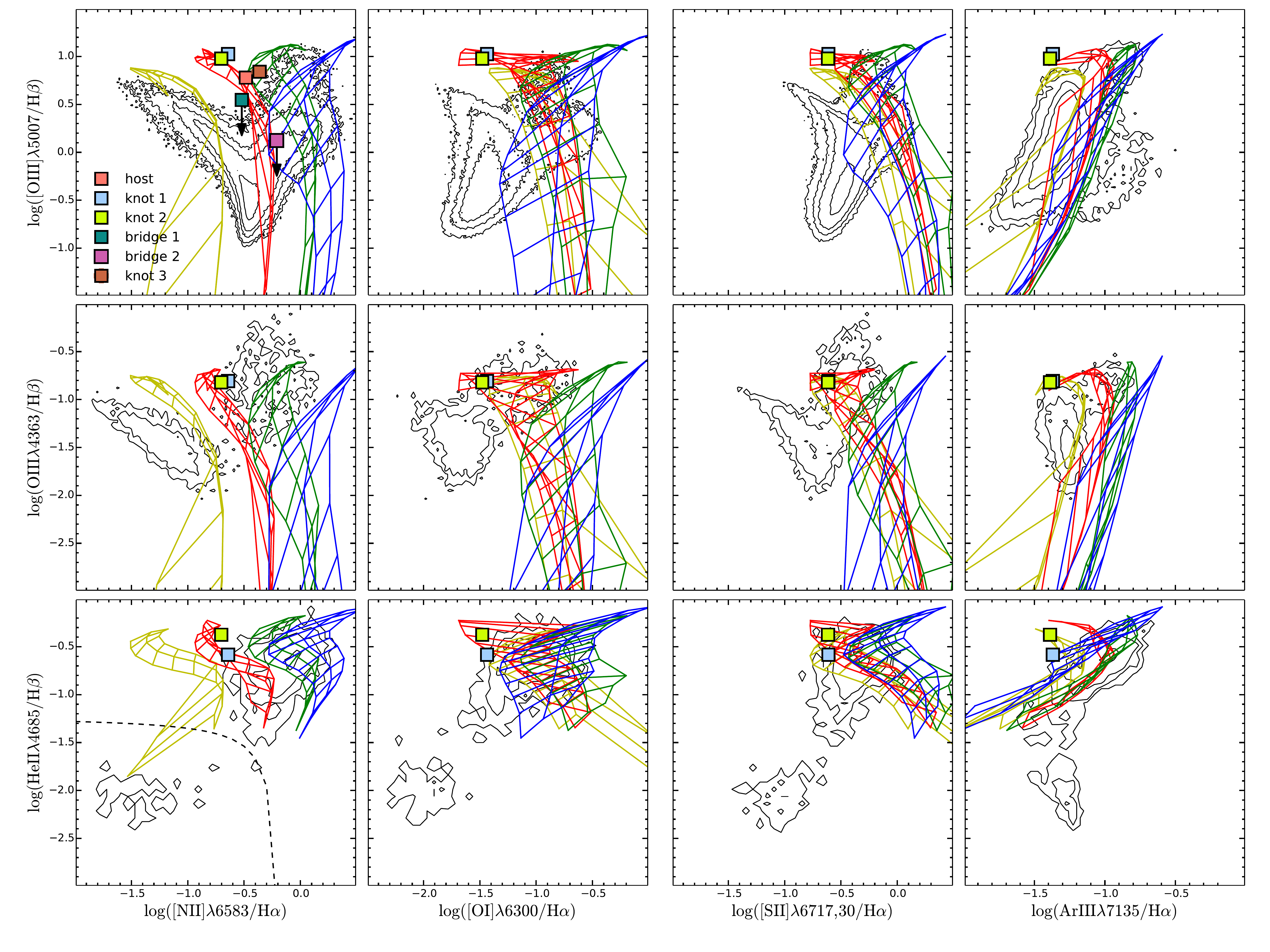}
 \caption{Various emission-line diagnostic diagrams for the extended emission-line region. The contours correspond to emission-line galaxies from the SDSS DR7 
catalogue \citep{Brinchmann:2004}, including 2000 galaxies with He\,II line measurements \citep{Shirazi:2012} as a reference for typical line ratios among galaxies. We only consider galaxies with a 
S/N$>$5 in each line for a given diagnostic plot. We compare the corresponding line ratios for the ENLR around PG~1307+085 (squared symbols) with a grid of dusty radiation 
pressure-dominated photoionization models \citep{Groves:2004}. We show models with four different metallicties 0.5$Z_\odot$ (yellow lines), 1$Z_\odot$ (red lines), 2$Z_\odot$ (green lines), 
4$Z_\odot$ (blue lines) and three power-law slopes for the ionizing continuum $\alpha=-1.2,-1.4,-1.7$ covering a large range in ionization parameters $0<\log(U)<-4$.}
 \label{fig:BPT}
\end{figure*}

\subsection{Physical conditions across the ENLR}
In Fig.~\ref{fig:spectra} we present the co-added QSO-subtracted spectra for the host galaxy, the three knots across the ENLR, and the 
bridge split up into two regions. Several narrow emission lines are visible  in the spectra. Besides the strong lines [\OIII]\,$\lambda\lambda$4960,5007, \Hb, [\NII]\,$\lambda\lambda$6548,6583, 
and \Ha, we  detect fainter lines such as the [\OIII]\,$\lambda4363$, \Hg, \HeII\,$\lambda4635$, [\SII]\,$\lambda\lambda6713,6730$, and [\ArIII]\,$\lambda7135$ only in the two brightest knots. 
We model the lines above a local continuum level with Gaussians profiles that are all coupled in their radial velocity and intrinsic velocity dispersion. The obtained line fluxes and kinematic 
parameters are listed in Table~\ref{tbl:line_measurements};  $3\sigma$ upper limits are given for the undetected lines based on the noise in the continuum. 

\begin{table*}\centering
\begin{footnotesize}
\caption{Emission-line fluxes and physical conditions for the narrow-emission lines in the QSO and extended region spectra}
\label{tbl:line_measurements}
\begin{tabular}{ccccccccc}\hline\noalign{\smallskip}
 & \multicolumn{2}{c}{QSO}  & host & knot 1 & knot 2 & knot 3 & bridge 1 & bridge 2 \\
 & 1st component & 2nd component  & & & & & \\\hline\smallskip
$\Delta v_{\mathrm{host}}$ [$\mathrm{km\,s}^{-1}$] & $120\pm12$ & $59\pm12$ & $67\pm13$ & $-84\pm12$ & $-17\pm12$ & $-101\pm13$ & $-38\pm27$ & $-246\pm28$\\
$\sigma$ [$\mathrm{km\,s}^{-1}$] & $110\pm7$ & $285\pm7$ & $122\pm7$  & $31\pm15$ & $74\pm7$ & $46\pm9$ & $106\pm26$ & $100\pm19$\\
H$\gamma$\,${}^{1}$ & $13.72\pm0.00$ & $17.83\pm0.73$  & $<2.53$ & $3.39\pm0.13$ & $2.96\pm0.08$ & $<0.23$ & $<0.25$ & $<0.49$\\
$[\mathrm{OIII}]\,\lambda4364$\,${}^{1}$ & $14.78\pm0.74$ & $17.60\pm1.73$ & $<2.15$ & $1.25\pm0.12$ & $1.02\pm0.08$ & $<0.28$ & $<0.27$ & $<0.58$\\
HeII\,$\lambda4685$\,${}^{1}$ & $6.31\pm0.21$ & $13.15\pm0.48$ & $<1.93$ & $2.10\pm0.11$ & $2.84\pm0.07$ & $<0.19$ & $<0.22$ & $<0.45$\\
H$\beta$\,${}^{1}$ & $32.39\pm1.12$ & $41.42\pm2.30$ & $9.20\pm0.72$ & $8.02\pm0.10$ & $6.72\pm0.07$ & $0.50\pm0.07$ & $<0.21$ & $<0.47$ \\
$[\mathrm{OIII}]\,\lambda5007$\,${}^{1}$ & $334.08\pm8.12$ &$414.68\pm9.92$  & $55.38\pm0.90$ & $85.02\pm0.24$ & $63.92\pm0.08$ & $3.45\pm0.10$ & $1.21\pm0.10$ & $1.04\pm0.16$\\
OI\,$\lambda6300$\,${}^{1}$ & $7.94\pm0.25$ & ... & $<2.59$ & $0.89\pm0.08$ & $0.69\pm0.06$ & $<0.15$ & $<0.21$ & $<0.31$\\
H$\alpha$\,${}^{1}$ & $28.94\pm0.72$ & $333.72\pm1.07$ & $31.12\pm0.64$ & $24.28\pm0.12$ & $20.82\pm0.08$ & $1.47\pm0.08$ & $0.56\pm0.10$ & $0.97\pm0.18$\\
$[\mathrm{NII}]\,\lambda6583$\,${}^{1}$ & $23.04\pm0.81$ & $98.71\pm1.57$ & $10.20\pm1.14$ & $5.54\pm0.16$ & $4.13\pm0.13$ & $0.64\pm0.08$ & $0.17\pm0.16$ & $0.60\pm0.12$\\
$[\mathrm{SII}]\,\lambda6717$\,${}^{1}$ & $10.03\pm0.35$ & ... & $<2.22$ & $3.39\pm0.28$ & $2.92\pm0.08$ & $<0.47$  & $<0.35$  & $<0.37$\\
$[\mathrm{SII}]\,\lambda6730$\,${}^{1}$ & $10.17\pm0.07$ & ... & $<2.45$ & $2.55\pm0.07$ & $2.15\pm0.06$ & $<0.14$ & $<0.20$ & $<0.41$\\
ArIII\,$\lambda7135$\,${}^{1}$ & $8.21\pm0.31$ & ... &  $<1.68$ & $1.04\pm0.06$ & $0.85\pm0.05$ & $<0.14$ & $<0.17$ & $<0.30$\\
$\langle\Sigma_{\mathrm{[OIII]}\lambda5007}\rangle$\,${}^{2}$ & ... & ... & $12.8$ & $9.0$ & $3.8$ & $0.2$ & $0.1$ & $0.2$ \\
$E(B-V)$ & ... & ... & $0.17\pm0.04$ & $0.06\pm0.01$ & $0.08\pm0.01$ & $0.04\pm0.12$ & ... & ... \\
$n_e$ [$\mathrm{cm}^{-2}$]  & $748\pm98$ & ...& ...& $85\pm133$ & $60\pm20$ & ... & ... & ... \\
$T_e$ [K] & $26118\pm1467$ & ... &  ... & $13437\pm578$ & $13959\pm259$ & ... & ... & ... \\
$\log(M_\mathrm{ion}/[M_\odot])$ & \multicolumn{2}{c}{$7.1$} & $7.0$ & $6.8$ & $6.9$ & $5.7$ & $5.3$ & $5.5$\\
\noalign{\smallskip}\hline\end{tabular}

\begin{flushleft}
${}^{1}$ Emission-line fluxes are given in units of $10^{-16}~\mathrm{erg}\,\mathrm{s}^{-1}\,\mathrm{cm}^{-2}$\\
${}^{2}$ Emission-line surface brightness is  given in units of $10^{-19}~\mathrm{erg}\,\mathrm{s}^{-1}\,\mathrm{cm}^{-2}\,\mathrm{arcsec}^{-2}$
\end{flushleft}
\end{footnotesize}
\end{table*}

From the \Ha/\Hb\ line ratio we infer the line of sight extinction $E(B-V)$ adopting a theoretical Balmer decrement of \Ha/\Hb=2.85 for case B recombination and a Milky 
Way like extinction curve \citep{Cardelli:1989}. Subsequently, we correct the observed line fluxes for extinction and use the [\SII]$\lambda 6730$/[\SII]$\lambda 6716$ line ratio to measure 
the electron density ($n_e$) and the ([\OIII]$\lambda 5007$+[\OIII]$\lambda 4960$)/[\OIII]$\lambda 4636$  ratio to estimate the electron temperature ($T_e$) following the prescription of 
\citet{Osterbrock:2006} as implemented in the \textsc{PyNeb} package \citep{Luridiana:2015}. We obtain values of $n_e\sim 100~\mathrm{cm}^{-2}$ and $T_e\sim 14000~\mathrm{K}$ for the two brightest 
knots of the ENLR as well as  $n_e\sim 1000~\mathrm{cm}^{-2}$ and $T_e\sim 25000~\mathrm{K}$ for the unresolved emission of the QSO. Our measurements for knot 1 are consistent 
with the results from the long-slit spectrum aligned along that region as reported by \citet{Oh:2013}, but with significantly smaller error bars. The measured temperatures and densities are 
typical conditions for the nucleus \citep[e.g.][]{Vaona:2012} and for the ENLR on kpc scales \citep[e.g.][]{Bennert:2006a,Bennert:2006b}, respectively.

The total ionized gas mass is mainly set by the amount of ionized hydrogen. This can be easily inferred from the photons emitted by the Balmer recombination lines. Following \citet{Osterbrock:2006} 
the ionized gas mass can be approximated from the \Ha\ luminosity as
\begin{equation*}
M_\mathrm{ion}=\frac{1.4m_\mathrm{p}}{n_\mathrm{e}\alpha_{\mathrm{H}\alpha}^\mathrm{eff}h\nu_{\mathrm{H}\alpha}}L_{\mathrm{H}\alpha}=3.4\times 10^6\left(\frac{100\mathrm{cm}^{-2}}{n_\mathrm{e}}
\right)\left(\frac {L _{\mathrm{H}\alpha}}{10^{41}~\mathrm{erg\,s}^{-1}} \right)\mathrm{M}_\odot
\end{equation*}
where $m_\mathrm{p}$ is the proton mass, $n_\mathrm{e}$ is the electron density and $h$ is the Planck constant. Here, we compute the ionized gas mass separately for each region based on the 
attenuation correct \Ha\ luminosity and electron density. If the attenuation or electron density is not measurable for a given region, we adopt $E(B-V)=0.1$ and $n_e=70\mathrm{cm}^{-2}$, which are 
representative values from all other regions. Adding the ionized gas mass of all regions we infer a total ionized gas mass of the ENLR of approximately $M_\mathrm{ion}\sim2.5\times10^7~M_\odot$ 
which is uncertain at an order or magnitude level.

The two brightest knots of the ENLR show a large number of lines that we can detect with high confidence. In Fig.~\ref{fig:BPT} we show twelve different line ratio
plots including the well-known [\OIII]\,$\lambda5007$/\Hb\ vs. [\NII]\,$\lambda$6583/\Ha\ BPT diagram \citep{Baldwin:1981, Veilleux:1987}. Except for the line 
ratios in the bridge, all regions clearly indicate photoionization by the AGN \citep[e.g.][]{Kewley:2006,Stasinska:2006}. However, the [\NII]\,$\lambda6583$/\Ha, 
[\SII]\,$\lambda6713,30$/\Ha, [\OI]\,$\lambda6300$/\Ha\ and [\ArIII]\,$\lambda7135$/\Ha\ ratios in those regions are much smaller than the typical line ratio of obscured AGN. This is 
also the case for the line ratios in the spatially unresolved NLR of the QSO. Only a very small fraction of obscured AGN in SDSS occupy this line ratio space as reported by 
\citet{Groves:2006}. They suggest that this is caused by a lower metallicity of the ionized gas compared to the majority of AGN. On the other hand, \citet{Stern:2013} studied the line ratio for a 
sample of unobscured AGN, such as PG~1307$+$085, and found a significant trend with AGN luminosity for most of the line ratios, arguing for selection effects rather than a metallicity difference. 
Here, we note that in particular the low $\log$([\OI]/\Ha)$\sim$-1.5 line ratio is also at the very extreme end of the high-luminosity unobscured AGN population studied by 
\citet{Stern:2013}.

To roughly quantify the ionization properties of the gas, we compare all line ratios for knots 1 and 2 with dusty radiation pressure dominated photoionization model \citep{Groves:2004} in 
Fig.~\ref{fig:BPT} for four metallicites ($0.5Z_\odot$, $1Z_\odot$, $2Z_\odot$ and $4Z_\odot$), three power-law slopes of the ionizing AGN continuum ($\alpha=1.2,\,1.4,\,1.7)$, and a single gas 
density of $n=100~\mathrm{cm}^{-3}$ as implied by the [\SII] doublet ratio. 
Except for the [\ArIII]\,$\lambda7135$/\Ha\ ratios, the model with $Z=1Z\odot$, $\alpha=-1.2$ and $\log U=-1.3$ is closest to all line ratios within $<$0.2~dex. It 
is interesting that the [\ArIII]\,$\lambda7135$/\Ha\ line ratio is even lower than predicted by this ``best-fitting'' model. Since Argon and Helium are not depleted on to dust, Argon is a good 
metallicity indicator independent of the assumed dust composition but depends on ionization parameter. In our case, \HeII\,$\lambda5685$ implies a high ionization parameter independent of 
metallicity while the low  [\ArIII]\,$\lambda7135$/\Hg\ line ratio indicates significantly lower metallicites than in the majority of AGN.  

We can definitely rule out  shock or  shock+precursor ionization because the high [\OIII]/\Hb\ line ratio would require high excitation by fast shocks with $v_s>500$~\kms\ 
\citep{Allen:2008}. This is inconsistent with the observed quiescent kinematics of the gas (see next section) across the ENLR.

\begin{figure*}
\includegraphics[width=\textwidth]{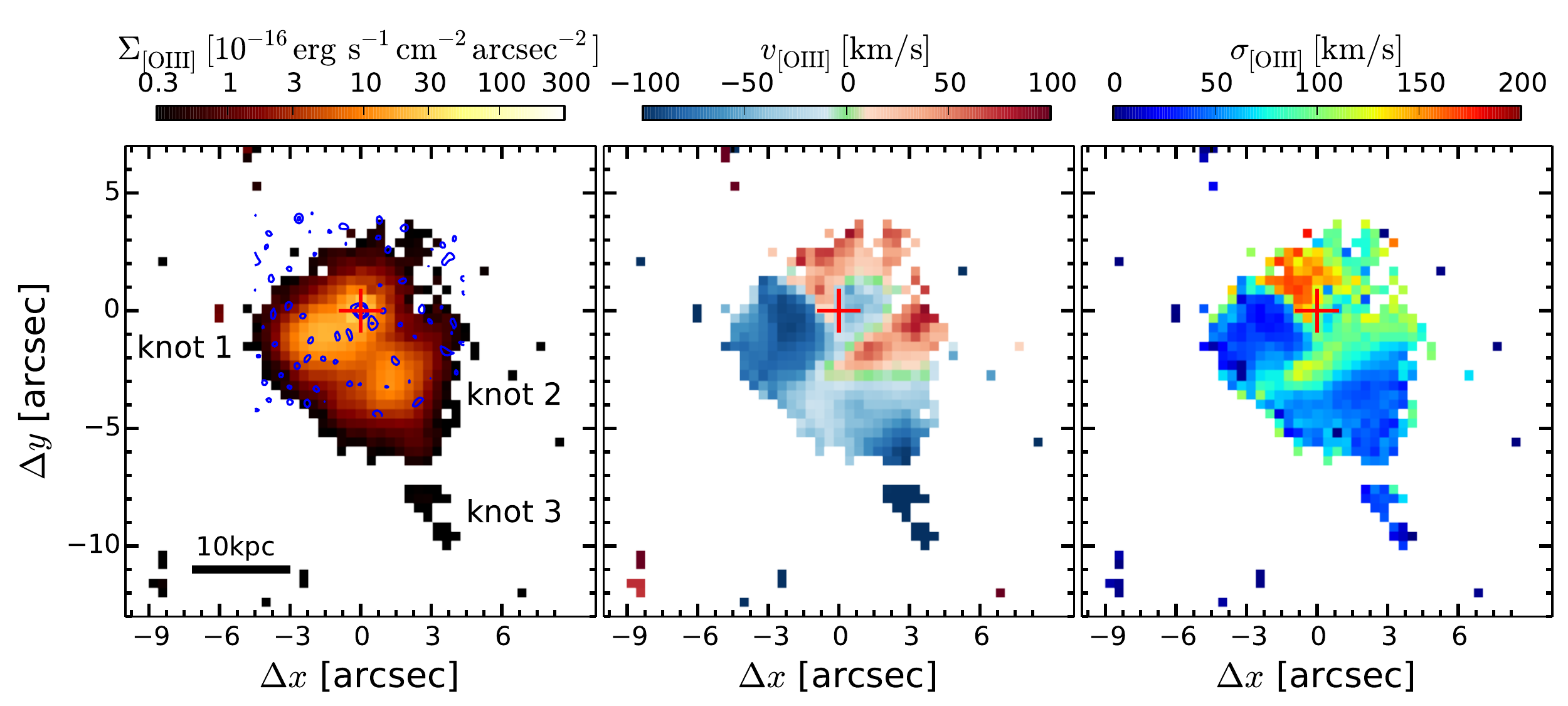}
\caption{Results from the spaxel-by-spaxel fitting of the [\OIII]$\,\lambda\lambda4960,5007$ emission lines after subtracting the bright QSO emission. \textit{Left panel:} 
[\OIII]$\,\lambda5007$ surface brightness map with a logarithmic colour scaling. The 5~GHz VLA radio map with a beam size of 0.4~arcsec is over-plotted as blue contours. Apart from the compact 
radio source at the nucleus, some weak extended radio emission is detected $\sim$1~arcsec towards the south-west in the direction of knot 2. All the other radio peaks are consistent with noise. 
\textit{Middle panel:} Radial velocity maps 
of the ionized gas with respect to the systemic redshift of the 
host galaxy. \textit{Right panel:} Intrinsic velocity dispersion map of the ionized gas corrected for instrumental resolution and redshift broadening. The peak 
position of the subtracted QSO is denoted by the red cross in all three panels.}
 \label{fig:O3_maps}
\end{figure*}

\subsection{Ionized gas kinematics}
In addition to the line measurements from the co-added spectra for the specific regions, we reconstruct the kinematics of the ENLR from the bright [\OIII]\,$\lambda\lambda4960,5007\AA$ 
 lines across the field. We model the doublet line independently for each spaxel as a single pair of Gaussians that are coupled in redshift and velocity dispersion with a fixed flux ratio of 
[\OIII]\,$\lambda5007$/[\OIII]\,$\lambda4960=3$ in the QSO-subtracted data cube. The best-fitting [\OIII]\,$\lambda5007$ surface brightness, radial velocity (systemic velocity subtracted) and 
velocity 
dispersion maps are shown in Fig.~\ref{fig:O3_maps}. Although the [\OIII]\,$\lambda5007$ in the QSO spectrum exhibits a strong blue-shifted component with 
a line width of  $\sim650$~\kms\ (FWHM), we detect mainly narrow lines with $<120$~\kms (FWHM) across the entire region. 

An increased line dispersion is apparently seen in the region between knot 1 and knot 2. The increase in velocity dispersion is not physical, but rather a combined artefact of beam smearing and 
fitting only a single Gaussian component. We noticed that the spectra in this region appear double peaked because the emission lines of knot 1 and knot 2 are offset by about 50\,km/s in velocity 
space and both components contribute with an almost equal surface brightness to the overall line shape. The same superposition also happens in the host galaxy region which is contaminated by 
emission from knot 1 due to beam smearing. The spectral resolution of MUSE is not sufficient to reliably decompose those lines in their respective components. Whether the same mechanism also 
leads to the higher velocity dispersion  $\sim$70~\kms\ in the region between the host galaxy and knot 2 is less clear. The higher than average dispersion may also be caused by weak shocks or 
dispersion-dominated motion in the gravitational potential of the host galaxy.  

The line-of-sight radial velocity variation shows a low amplitude of $<\pm200$~\kms\ in general. Interestingly, we find a smooth velocity gradient from red-shifted to significantly blue-shifted with 
respect to the systemic velocity from the location of the host galaxy through knot 2 towards knot 3.  This velocity gradient has the largest velocity amplitude of 200~\kms. We consider this 
gradient physical and unrelated to instrumental effects as beam smearing would only smooth out the gradient. From the data alone we cannot distinguish between gas approaching and escaping from the 
host galaxy. 

The S/N in the bridge between PG~1307$+$085 and the massive galaxy at the group centre is too low for analysing individual spaxels. However, the radial velocity measured from 
the two binned regions along the bridge (see Table~\ref{tbl:line_measurements}) reveals that the gas is more blue-shifted closer to the massive companion by $\sim$260~\kms. This is consistent in 
sign and amplitude with the redshift difference between two galaxies. It confirms that the ionized gas bridge is not just a superposition, but physically connects the ionized gas between the two 
galaxies. 

\section{Discussion}
\subsection{The BH mass - $\sigma_*$ relation}
\begin{figure}
 \resizebox{\hsize}{!}{\includegraphics{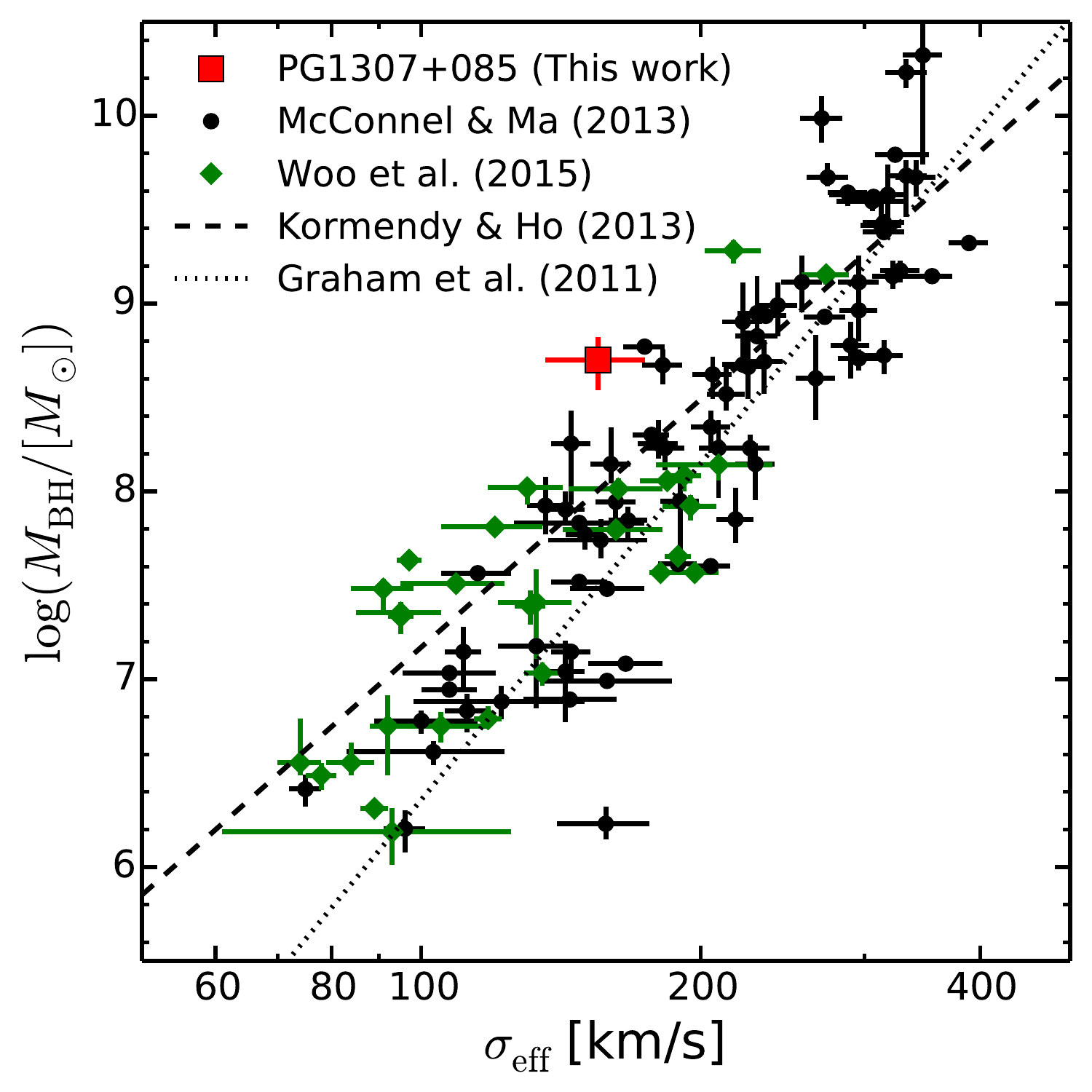}}
 \caption{\mbh-\s~relation for a sample of quiescent local galaxies \citep[black points;][]{McConnell:2013} and reverberation-mapped AGN
\citep[green symbols;][]{Woo:2015}. For comparison we also plot the relation of \citet{Kormendy:2013} as a dashed line and \citet{Graham:2011} as a dotted line. The position of PG1307+085 
is shown as a red squared point adopting the reverberation-mapped BH mass inferred by \citet{Peterson:2004}.}
 \label{fig:MBH_sigma}
\end{figure}

With the unique sensitivity, spatial resolution and image quality of MUSE, we are able to measure a robust $\sigma_*$ for the faint host galaxy underneath the luminous QSO PG~1307+085. Here we 
compare our measurements with the $M_\mathrm{BH}$--$\sigma_*$ relation of local quiescent galaxies \citep[][72 objects]{McConnell:2013}, where BH masses are derived directly from 
dynamical measurements, as well as reverberation-mapped AGN \citep[][29 objects]{Woo:2015}. We also show the best-fitting relation derived by \citet{Kormendy:2013} (51 objects; pseudo bulges and 
mergers excluded) and by \citet{Graham:2011} (64 objects; all morphologies). As can be seen from Figure~\ref{fig:MBH_sigma}, PG~1307+085 is consistent with the local 
relation within its intrinsic scatter taking the errors of our measurements into account. However, it is slightly offset towards a higher BH mass with respect to the mean relation. 

An offset of $\Delta M_\mathrm{BH}(@ z=0.15)=0.24$~dex is suggested by the reported redshift evolution of \citet{Woo:2006}. The nominal offset from the mean relation we measure for 
PG~1307$+$085 is about 0.3~dex and would be in line with the predicted offset. For a single object it is, however, still within the intrinsic scatter of the relation. Certainly, the offset from the 
relation of quiescent galaxies is smaller than in some extreme AGN host galaxies \citep{Husemann:2011, Trakhtenbrot:2015} and can be accounted for by the uncertainties in the $M_\mathrm{BH}$ 
and $\sigma_*$ estimation. Thus, we consider the offset insignificant given the systematic uncertainties for a single object. In case that the offset is real, it 
is possible that either PG~1307$+$085 is moving further away from the relation given the substantial BH growth compared to the low star formation in the spheroid implied by the FIR colours, or that 
the spheroid growth will eventually catch up the BH growth to move it back towards the relations.

Given that we only consider a measurement for a single object, we refrain from discussing any further physical interpretation here. Instead, we highlight the efficiency and potential of 
measuring $\sigma_*$ of luminous QSOs with MUSE in the context of previous studies. PG~1307$+$085 is the third most massive BH and most luminous AGN among the 
reverberation-mapped AGN with $\sigma_*$  measurements discussed by \citet{Woo:2015}. Measurements of $\sigma_*$ for luminous QSOs are difficult because of the high AGN-to-host/spheroid contrast 
ratio. 
Different techniques have been used to tackle this problem. Since the contrast ratio is much weaker in the NIR, NIR observations, in particular with adaptive optics (AO), can much better resolve the 
stellar continuum against AGN contribution. Stellar velocity dispersions have been obtained for about a dozen QSOs at $z<0.1$ \citep{Dasyra:2007,Watson:2008,Woo:2010,Woo:2013}. However, AO-assisted 
NIR observations for higher redshift QSOs ($z>0.1$) are time consuming and may suffer from a low S/N \citep{Grier:2013} caused by the combination of weak absorption lines, surface brightness dimming, 
instrument sensitivity and atmospheric transmission. 

In the rest-frame optical regime, the QSO and host galaxy contribution can be separated in high S/N single-aperture spectra by directly modelling them with template spectra for each component 
\citep{Shen:2008,Shen:2015}. While a large sample can be studied with this technique, the fixed aperture size prevents such studies from measuring $\sigma_*$ at a physical size 
matched to the spheroid component. Furthermore, the reliability of the measurements decreases with increasing AGN luminosity and contrast ratio, thus limiting this technique to AGN with 
$L_{5100}\lesssim 
10^{45}~\mathrm{erg\,s}^{-1}$. To probe more luminous AGN, deep off-axis spectroscopy a few arcsec away from the QSO has been obtained to reduce the contrast between QSO and host galaxy emission 
\citep{Hughes:2000,Wolf:2008,Wold:2010}.  While $\sigma_*$ can be reliably derived from these off-axis spectra \citep{Wolf:2008}, an uncertain aperture correction needs to be applied to infer 
comparable measurements representative for the spheroidal component. 

A direct decomposition the stellar continuum of the host galaxy underneath very luminous QSOs with deep optical IFU observations was previously attempted with VIMOS for 
HE~1029$-$1401 \citep{Husemann:2010}. Although HE~1029$-$1401 is much closer with a redshift of $z=0.086$ but has a similar luminosity as PG~1307$+$085, the achieved S/N in the host spectrum was 
significantly lower than what we obtained with MUSE for PG~1307$+$085. This can be clearly attributed to the superior sensitivity, spatial resolution/sampling, and image quality of MUSE compared to 
VIMOS. These capabilities allow direct measurements of $\sigma_*$ with MUSE even in a high contrast regime that has not been accessible to any other techniques so far.

\subsection{The nature of the large extended emission line region}
\subsubsection{Ionized gas shells driven by AGN outflows?}
A number of recent studies report kpc-scale high-velocity outflows to be common around luminous AGN with line widths of about 1000~\kms\
\citep[e.g.,][]{CanoDiaz:2012,Liu:2013b,Liu:2014,Harrison:2014}. Those outflows have been interpreted as spherically symmetric radiation pressure dominated winds \citep{Liu:2013b} and are possibly 
common for AGN with $L_\mathrm{bol}>2\times10^{45}~\mathrm{erg\,s}^{-1}$ \citep{Zakamska:2014}. However, not all QSOs above this luminosity threshold exhibit such powerful 
superbubble-like outflows \citep[e.g.,][]{Husemann:2010,Husemann:2013a}. The ionized gas kinematics of the ENLR around PG~1307$+$085 exhibits only small line widths with a velocity dispersion of 
$<$50~\kms\ in all the bright extended knots without any asymmetric component. This confidently rules out the AGN-driven super-bubble scenario at least for this particular QSO. 

Although PG~1307$+$085 is classified as radio-quiet QSO with $R<0.1$ \citep{Kellermann:1989} and a relatively low radio-power ($L_\mathrm{radio}<2\times10^{23}~\mathrm{W\,Hz}^{-1}$), it is still  
possible that a small-scale ($<$1~kpc) radio jet is accelerating gas through jet-cloud interactions. Some evidence that there is an outflow at least in the circumnuclear region of 
PG~1307$+$085 comes from the asymmetric [\OIII] line  \citep[e.g.,][]{Mullaney:2013}. A deep 4.8~GHz high-resolution map obtained with the VLA was presented by \citet{Leipski:2006b}. They reported 
tentative but inconclusive evidence for a small-scale 
one-sided radio jet with a size of 1~arcsec towards the south-west direction. The jet is certainly much smaller than the ENLR in size and cannot be the driver of its morphology. Especially, it 
would not be able to explain the bright ENLR knot to the east almost 90$^{\circ}$ away from the putative jet axis. On the other hand, the direction of the radio jet is consistent with the 
orientation of the velocity gradient close to the QSO position where we also find a peak in ionized gas velocity dispersion. 

To summarize, we cannot rule out that there is an AGN outflow on scales of $\ll$1~arcsec, potentially driven by a radio jet of the same size. This picture is in agreement with the correlation 
between [\OIII] asymmetry and radio luminosity discussed by \citet{Mullaney:2013}, because the low velocity dispersion of the blue-shifted [\OIII] component is consistent with the low $R$ parameter 
and radio power. Another possibility is that there was a past outflow event, either driven by the AGN or radio jet. Such an event could have lifted gas further out in the halo of the galaxy 
(without accelerating the gas to escape velocity) from which the ENLR has formed.

\subsubsection{Galaxy interactions and tidal debris?}
Galaxy interactions, in particular major mergers, have been thought as a common channel to trigger luminous AGN \citep[e.g.][]{Sanders:1988a,Hopkins:2005}. While a 
fraction of AGN are known to reside in ongoing major mergers, systematic studies of QSO hosts have shown that major mergers cannot be the dominant mechanisms for fueling AGN 
\citep[e.g.,][]{Cisternas:2011,Villforth:2014} at least for moderate luminosities. However, it is particularly difficult to identify the faint morphological disturbances and tidal features 
in advanced mergers or major merger remnants, which can only be seen in very deep exposures \citep[e.g.,][]{Bennert:2008,Duc:2015}. Indeed, large ENLRs have been observed that are clearly associated 
with ongoing galaxy interaction \citep[e.g.,][]{Villar-Martin:2010,daSilva:2011}.

In the case of PG~1307$+$085, neither the \textit{HST} images \citep{Bahcall:1994,Veilleux:2009}, nor the deep AO-assisted ground-based image \citep{Guyon:2006} reveal any detectable 
substructure at the location of the bright knots in the ENLR. However, the amount of ionized gas in these regions is small ($\sim10^7M_\odot$) and given the hard radiation field, the total gas mass 
is not likely to be more than an order of magnitude larger. Thus, we cannot rule out that the ENLR corresponds to tidal features of a merging event. However, the likelihood of a recent major merger 
is low given that PG1307$+$085 is known to be FIR faint \citep{Netzer:2007}, implying a lower star formation rate - at least in the last 100 Myr 
\citep{Hayward:2014} - than in other QSOs of similar luminosity. While the time delay between the peak of star formation and the AGN phase during the merger of a few 100\,Myr could provide an 
explanation \citep[e.g.,][]{Wild:2010}, a minor merger scenario is more likely, considering the low gas mass and  the lower metallicity gas. In particular knot 2 and the diffuse extension towards 
knot 3 with the strong velocity gradient may be a kinematic signature of the tidal debris from a minor companion.  

Such a minor merger scenario cannot explain the bright knots 1 and 2 in the ENLR morphology at the same time because they are oriented almost exactly 90$^\circ$ with respect to their 
specific QSO line-of-sights. It is notable that knot 1 is positioned along the axis between the massive galaxy at the galaxy group's centre and the QSO which are connected by a 
very faint bridge of ionized gas. Thus, the ENLR may be linked to the vivid  group environment leading to gravitational interaction with several major or minor galaxies over short periods of time. 

\subsubsection{Illumination or interactions with the inter-group medium?}
Another possibility for the origin of the ENLR morphology around PG~1307$+$085 is an ongoing interaction of the host galaxy with the diffuse inter-group medium (IGM). The systemic 
redshift of the host galaxy differs from the redshift of the central group galaxy by $\sim$350~\kms\ implying that it moves with significant speed through the IGM. 
Ionized gas tails behind galaxies due to ram-pressure stripping of their gas are well-known phenomena in groups and clusters \citep[e.g.][]{Chung:2007,Fumagalli:2014} and large filamentary 
structures of  gas have been observed in cool-core clusters \citep[e.g.][ and reference therein]{Tremblay:2015}. Thus, it is possible that either ram pressure is pushing the remaining gas out of the 
QSO host or that the galaxy has hit a higher density IGM cloud/filament and is piling up the gas. 

In particular the (likely) low gas-phase metallicity together with the strong velocity gradient towards the south-west direction suggest an ongoing mixing of the gas with the surrounding IGM. The 
velocity gradient shows that the radial velocity of the most distant ionized gas clouds converges to the systematic velocity of the group. This would be common for both gas inflow and outflow 
scenarios because the gas will be confined by the ambient gas pressure anyway at large distances from the QSO. The gas phase metallicity cannot be used to discriminate between the 
two scenarios as it will be mixed with the IGM in both cases. Such a mixing is observed in the large ENLRs around powerful radio-loud AGN \citep[e.g.][]{Fu:2006} where the radio jet is likely 
driving a shock front through the IGM. 

While luminous AGN with low NLR gas-phase metallicites are rare \citep[e.g.][]{Groves:2006}, another QSO, HE~2158$-$0107, with very similar characteristics has been studied by 
\citet{Husemann:2011}. Similar to PG~1307$+$085, the ionized gas around that QSO exhibits a low [\NII]/\Ha\ ratio, implying low metallicities, and a large ENLR extending $>30$~kpc beyond the 
host galaxy. It is unclear whether the QSO is linked to a galaxy group such as PG~1307$+$085, but it is possible that the origin of the ENLR for both objects is the same. 
HE~2158$-$0107 appears to be more strongly offset from the $M_\mathrm{BH}$--spheroid relations towards higher BH masses. This may point either to a 
different formation mechanism or a different evolutionary phase in the process that is also responsible for producing their large ENLR. Currently, a detailed systematic study of ENLR around  
early-type/elliptical QSOs hosts is missing to test how common group interactions are among this specific population and what role those interactions have for the overall evolution of those galaxies.

\section{Summary}
In this article we present deep optical integral-field spectroscopy with MUSE of the luminous radio-quiet QSO PG~1307$+$085. The combination of high spatial resolution and a wide FoV obtained with 
MUSE is essential to simultaneously characterize the large-scale distribution and properties of the warm ionized gas and the compact elliptical host galaxy underneath the bright 
QSO. We study the properties of this luminous QSO in unprecedented detail. Our findings can be summarized as follows:
\begin{enumerate}
\item We resolve the underlying host galaxy and measure a stellar velocity dispersion of $\sigma_*=155\pm19$~\kms. This places PG~1307$+$085 still within the scatter of the local
$M_{\mathrm{BH}}$--$\sigma_*$ relation based on the reverberation-mapped BH mass, but slightly offset by $\sim$0.3~dex towards a higher BH with respect to the mean relation. It is unclear whether 
PG~1307$+$085 continues to move away from the mean relation or that the spheroid  growth will eventually overtake the BH growth again moving the system back towards the relation.
 \item We detect an ENLR with a complex morphology that extends at least out to $\gtrsim30$~kpc from the QSO. In addition, we also reach a very low surface 
brightness detection limit to identify an ionized gas bridge between PG~1307$+$085 and a more massive companion galaxy just about 50~kpc away, presumably at the centre of a galaxy group. 
 \item The kinematics of the ENLR reveals no obvious signatures of broad emission lines on kpc scales which would be a genuine signature for large-scale AGN-driven outflow. Such a quiescent 
kinematic does not agree with the expectation that luminous AGN with $L_{\mathrm{bol}}>10^{45.3}~\mathrm{erg\,s}^{-1}$ develop a radiatively-driven galactic scale outflow 
\citep{Zakamska:2014}. 
Since PG~1307$+$085 has a low radio power even compared to the radio-quiet QSO population, our data are more consistent with the scenario drawn by \citet{Mullaney:2013} in which galactic size 
radio jets are required and responsible for developing kpc-scale outflows. 
\item A detailed comparison of line ratios across the ENLR with photoionization models indicates that the metallicity of the gas is likely lower compared to the bulk of the AGN population. 
Considering also the ENLR morphology and kinematics of the gas, we argue that PG~1307$+$085 is possibly in a phase of significant interaction with  the group environment during which 
either the gas is stripped from the host galaxy by ram pressure or the host galaxy is replenished with external gas by minor mergers or intra-group gas filaments.
\end{enumerate}

Although we cannot draw conclusions for the overall QSO population, the MUSE observations of  PG~1307$+$085 provide unprecedented insights regarding the current state and evolution of its 
host galaxy. This demonstrates the great potential of MUSE for studying QSO host galaxies simultaneously with their immediate environments. The on-axis measurements of $\sigma_*$ for such 
a luminous QSO with a very compact host galaxy opens new perspectives for understanding the $M_\mathrm{BH}$--spheroid relation for AGN at the high BH mass end. In the future, observations of QSO host 
galaxies will greatly benefit in particular from the optical wide-field ground-layer adaptive optics mode of MUSE with which QSO studies can be extended to even higher redshifts and QSO luminosities.

\section*{Acknowledgements}
We thank the referee for very instructive comments and suggestions that improved the clarity of the manuscript.
We would like to thank Brent Groves and Stefanie Komossa for their help and valuable discussions about AGN photoionization models and their interpretation. These observations would not have been 
possible with enthusiasm and support of the staff at Paranal and the entire MUSE team, in particular Roland Bacon as the PI of the instrument. We thank Jakob Walcher for the 
continuous support for the development of {\tt PyParadise}. The AGN Fe template spectrum of I~Zw~1 was kindly provided by Todd Boroson.  VNB acknowledges assistance from the National Science 
Foundation (NSF) Research at Undergraduate Institutions (RUI) grant AST-1312296 and support for program number HST-AR-12625.11-A, provided by NASA through a grant from the Space Telescope Science 
Institute, which is operated by the Association of Universities for Research in Astronomy, Incorporated, under NASA contract NAS5-26555. JS acknowledges the European Research Council for the 
Advanced Grant Program number 267399-Momentum. JHW acknowledges support by the National Research Foundation of Korea to the Center for Galaxy Evolution Research (2010-0027910). 

This research has made use of the NASA/IPAC Extragalactic Database (NED) which is operated by the Jet Propulsion Laboratory, California Institute of Technology, under contract with the National 
Aeronautics and Space Administration. All Figures have been created using Matplotlib library for Python \citep{Hunter:2007}. This research made use of APLpy, an open-source plotting package for 
Python hosted at http://aplpy.github.com. We used the online software CosmoCalc \citep{Wright:2006}  for the computation of cosmological parameters.

\bibliographystyle{mnras}
\bibliography{references}

\end{document}